\documentclass[conference,10pt]{IEEEtran}
\IEEEoverridecommandlockouts
\usepackage{cite}
\usepackage{amsmath,amssymb,amsfonts}
\usepackage{algorithmic}
\usepackage{graphicx,physics,array}
\usepackage{textcomp}
\usepackage{xcolor,soul}
\def\BibTeX{{\rm B\kern-.05em{\sc i\kern-.025em b}\kern-.08em
    T\kern-.1667em\lower.7ex\hbox{E}\kern-.125emX}}

\begin{document}

\title{Quantum Error Correction For Dummies
}

\author{\IEEEauthorblockN{Avimita Chatterjee}
\IEEEauthorblockA{\textit{CSE Department} \\
\textit{Penn State University}\\
PA, USA\\
amc8313@psu.edu}
\and
\IEEEauthorblockN{Koustubh Phalak}
\IEEEauthorblockA{\textit{CSE Department} \\
\textit{Penn State University}\\
PA, USA\\
krp5448@psu.edu}
\and
\IEEEauthorblockN{Swaroop Ghosh}
\IEEEauthorblockA{\textit{School of EECS} \\
\textit{Penn State University}\\
PA, USA\\
szg212@psu.edu}
}

\maketitle

\begin{abstract}
In the current Noisy Intermediate Scale Quantum (NISQ) era of quantum computing, qubit technologies are prone to imperfections, giving rise to various errors such as gate errors, decoherence/dephasing, measurement errors, leakage, and crosstalk. These errors present challenges in achieving error-free computation within NISQ devices. A proposed solution to this issue is Quantum Error Correction (QEC), which aims to rectify the corrupted qubit state through a three-step process: (i) detection: identifying the presence of an error, (ii) decoding: pinpointing the location(s) of the affected qubit(s), and (iii) correction: restoring the faulty qubits to their original states. QEC is an expanding field of research that encompasses intricate concepts. In this paper, we aim to provide a comprehensive review of the historical context, current state, and future prospects of Quantum Error Correction, tailored to cater to computer scientists with limited familiarity with quantum physics and its associated mathematical concepts. In this work, we, (a) explain the foundational principles of QEC and explore existing Quantum Error Correction Codes (QECC) designed to correct errors in qubits, (b) explore the practicality of these QECCs concerning implementation and error correction quality, and (c) highlight the challenges associated with implementing QEC within the context of the current landscape of NISQ computers. 
\end{abstract}


\section{Introduction} \label{introduction}

In recent years, quantum computing has garnered substantial interest owing to its potential to revolutionize diverse industry sectors, including cybersecurity, pharmaceuticals, finance, and manufacturing \cite{bova2021commercial}. Quantum computers employ qubits for the representation and computation of information. Qubits harness quantum-mechanical properties, such as superposition, entanglement, and interference, which theoretically endow quantum computers with a speed advantage over classical algorithms and computing systems. Quantum computing algorithms have been employed in various fields, including quantum machine learning \cite{schuld2015introduction}, optimization \cite{li2020quantum} and quantum chemistry \cite{cao2019quantum, aspuru2005simulated}. Quantum computers are realized through a diverse range of qubit technologies, such as trapped ion qubits \cite{debnath2016demonstration, brandl2016cryogenic}, photonic qubits \cite{wang2016experimental, qiang2018large}, superconducting qubits \cite{chow2012universal, wendin2017quantum}, quantum dots qubits \cite{recher2010quantum} and many more. However, for all these technologies, it is a challenging task to entirely isolate qubits from external noise, making errors in quantum computers inevitable. Consequently, quantum computers necessitate some form of error correction.

\begin{figure*}
    \centering
    \includegraphics[width=\linewidth]{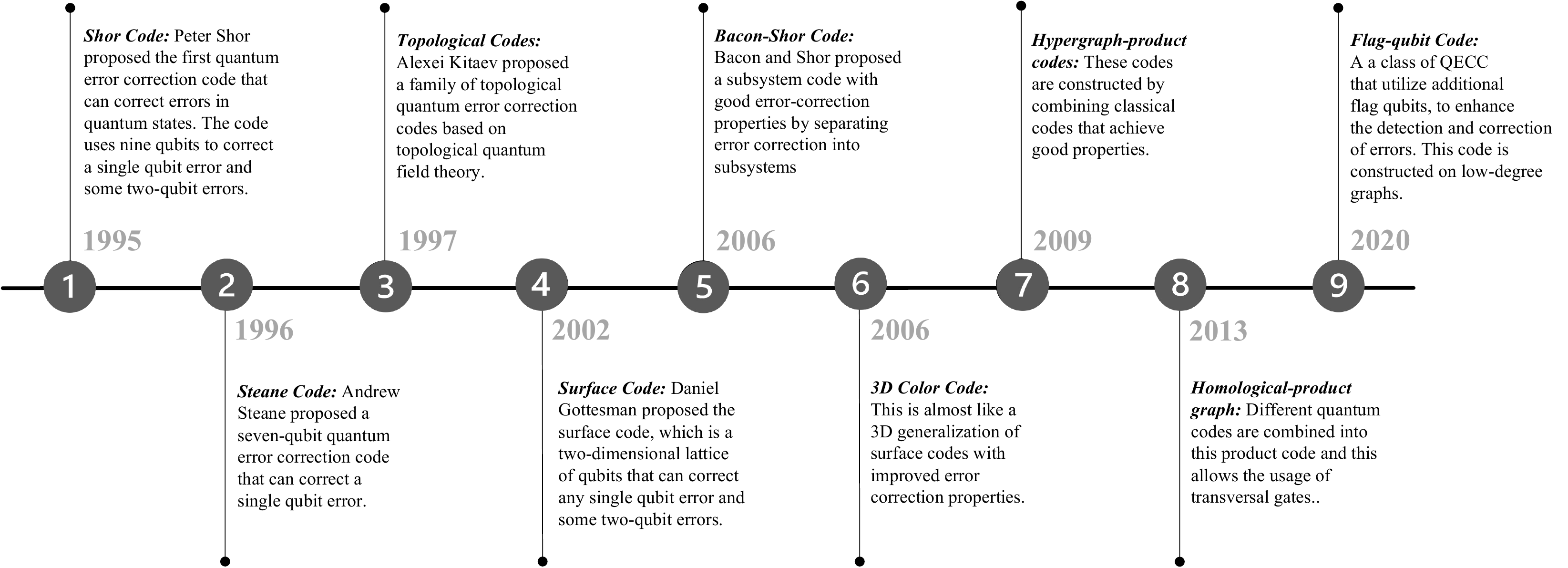}
    \caption{Advancement and Evolution of Quantum Error Correction. 
    \raisebox{.5pt}{\textcircled{\raisebox{-.9pt} {\textbf{1}}}}: \cite{shor1995scheme}, 
    \raisebox{.5pt}{\textcircled{\raisebox{-.9pt} {\textbf{2}}}}: \cite{steane1996simple},
    \raisebox{.5pt}{\textcircled{\raisebox{-.9pt} {\textbf{3}}}}: \cite{kitaev1997quantum}, 
    \raisebox{.5pt}{\textcircled{\raisebox{-.9pt} {\textbf{4}}}}: \cite{dennis2002topological}, 
    \raisebox{.5pt}{\textcircled{\raisebox{-.9pt} {\textbf{5}}}}: \cite{bacon2006operator}, 
    \raisebox{.5pt}{\textcircled{\raisebox{-.9pt} {\textbf{6}}}}: \cite{bombin2006topological}, 
    \raisebox{.5pt}{\textcircled{\raisebox{-.9pt} {\textbf{7}}}}: \cite{tillich2013quantum}, 
    \raisebox{.5pt}{\textcircled{\raisebox{-.9pt} {\textbf{8}}}}: \cite{haah2011local}, 
    \raisebox{.5pt}{\textcircled{\raisebox{-.9pt} {\textbf{9}}}}: \cite{chamberland2020topological}.
    }
    \label{fig:qec_timeline}
    \vspace{-5mm}
\end{figure*}

The established classical error correction theories have resulted in a high error tolerance for classical computers \cite{hamming1950error}. However, the adaptation of existing classical error correction techniques for quantum computing is challenging due to the no-cloning theorem, which prohibits the duplication of qubits similar to classical bits \cite{wootters1982single}. In addition, the measurement of qubits is subject to limitations as the act of measuring a qubit leads to the collapse of its wavefunction \cite{von2018mathematical}, resulting in the loss of its quantum state. The year 1995 saw the proposal of the first Quantum Error Correction (QEC) scheme by Peter Shor \cite{shor1995scheme}. 
Fig.\ref{fig:qec_timeline} illustrates the chronology of significant advancements in the field of quantum error correction codes. For a comprehensive comparison of these significant developments, please refer to the Table \ref{tab:QECC_comps}.

Multiple review articles exist on QEC \cite{roffe2019quantum, devitt2013quantum, matsumoto2021survey} and associated topics \cite{preskill1998reliable, gaitan2008quantum, lidar2013quantum}. Nonetheless, these articles can be difficult to grasp, as they often involve complex mathematical concepts and implicitly assume that readers possess pre-existing knowledge about the domain. The aim of this work is to provide a comprehensive yet accessible introduction to the fundamental concepts of quantum error correction for researchers who may not have an extensive background in quantum physics or related mathematical fields. It is not necessary for readers to have prior knowledge of Quantum Error Correction (QEC) for this review. However, it is assumed that the readers are familiar with quantum circuit notations as described in \cite{nielsen2002quantum}, including basic measurement operations, controlled-NOT gate $(CNOT)$, and Hadamard gate $(H)$. \footnote{As a notational shorthand, we often remove the tensor product sign, $\otimes$ when we denote the tensor product of multiple operators. For example, $X \otimes Y \otimes Z$ may be interchangeably written as $XYZ$.}

We commence by introducing the foundational concepts of quantum computing in Section \ref{prelim}, including qubits, quantum gates, quantum circuits, types of errors, and the distinctions between classical and quantum error correction. Section \ref{qec_fundamentals} elucidates the essential principles of QEC, beginning with repetition codes and progressing to stabilizer formalism and topological codes. Sections \ref{qec_application} and \ref{qec_future} address the practical applications of Quantum Error Correction Codes (QECC) and the challenges associated with their future development, respectively.
\section{Preliminaries} \label{prelim}
\subsection{An Overview of Quantum Computing}
\paragraph{\textbf{Qubits}} Qubits are fundamental units of a quantum computer that are analogous to classical bits. In general, a qubit is represented by a quantum state 
$\ket{\psi}=$
$\big[\begin{smallmatrix}
  \alpha\\
  \beta
\end{smallmatrix}\big]$, where $\alpha$ and $\beta$ are complex amplitudes such that $|\alpha|^2$ represents the probability of qubit being measured to classical 0 and $|\beta|^2$ represents the probability of qubit being measured to classical 1 i.e. $|\alpha|^2+|\beta|^2=1$. Every qubit has two fundamental basis states, $\ket{0}$ ($\alpha=1$, $\beta=0$) and $\ket{1}$ ($\alpha=0$, $\beta=1$). Qubit undergoes unitary gate operations that change its state and finally, a measurement operation is performed to collapse the qubit onto either 0 or 1 values. 

\paragraph{\textbf{Quantum gates}} Quantum gates are unitary matrix operations that operate on single or many qubits to change their states. They are realized using different methods based on the qubit technology such as using microwave pulses in superconducting qubits, laser pulses in trapped ion and quantum dots, and radio frequency pulses in Nuclear Magnetic Resonance (NMR) qubits. The gate operation speeds also vary with technologies, ranging from picoseconds (photonic qubits) to a few seconds (NMR qubits) \cite{ladd2010quantum}. Single-qubit gates, for example, include the X (NOT) gate, H (Hadamard) gate, and rotation gates such as $R_X$, $R_Y$, $R_Z$, and $U$ gate. Two-qubit gates include the CNOT (controlled-NOT) gate, Toffoli gate, controlled rotation gates, and Peres gate \cite{peres1985reversible}.

\paragraph{\textbf{Quantum circuit}} A quantum circuit is an ordered sequence of gate operations performed over time. A quantum circuit comprises of state initialization/preparation to prepare the initial state of the qubits, which are then transformed to the desired state using gate operations in the circuit and finally measured using a measurement gate. All quantum operations are performed in quantum Hilbert space \cite{von2018mathematical} and the high-level gates in the circuit e.g., Tofolli are broken down into a native gate set of the quantum hardware prior to execution. This process is referred to as transpilation. 

\subsection{Types of Errors in Quantum Computing} \label{qc_errors}

Noise in quantum computing  refers to any unwanted influence on qubits that leads to errors in the basis state. Primarily, there are two types of errors: bit-flip errors and phase-flip errors \cite{nielsen2002quantum}. Bit flip error, also known as $X$ error, occurs when the state of the qubit is flipped i.e. $\ket{0}$ changes to $\ket{1}$ and vice-versa. On the other hand phase flip error, also known as $Z$ error, involves the sign of the qubit's phase i.e. $\ket{1}$ changes to $-\ket{1}$ but $\ket{0}$ remains $\ket{0}$. To sum up if we have a basis state, $\ket{\psi} = \alpha\ket{0} + \beta\ket{1}$, then $X\ket{\psi} = \alpha\ket{1} + \beta\ket{0}$ and $Z\ket{\psi} = \alpha\ket{0} - \beta\ket{1}$. Both these errors can interact with each other and give rise to more complex errors in the system. A brief explanation of errors that can arise in the system is as follows:



\paragraph{\textbf{Gate error}} Gate error occurs when a quantum gate changes state of qubit(s) incorrectly. They are represented by fidelity, which denotes the probability of error in computation.

\paragraph{\textbf{Decoherence error}} Decoherence error occurs when a qubit interacts with the environment thereby losing its coherence and becoming an entangled state.  

\paragraph{\textbf{Measurement error}} Measurement error occurs when the classically measured output from a measurement operation is incorrect.

\paragraph{\textbf{Crosstalk error}} Crosstalk error occurs when a qubit interacts with a physically adjacent qubit, leading to an unwanted alteration to the qubit state.

\subsection{Classical and Quantum Error Correction}

\paragraph{\textbf{Classical Error Correction}} In classical computing, error correction is employed to maintain the integrity and precision of digital data by identifying and rectifying errors that may have arisen during transmission. This process utilizes Error Correction Codes (ECC) \cite{mackay2003information, hamming1950error}. 
The most commonly used ECC include Hamming codes \cite{hamming1950error}, Bose-Chaudhuri-Hocquenghem (BCH) codes \cite{bose1960class} and Reed-Solomon codes \cite{reed1960polynomial}. All of these codes detect and correct errors by adding redundant information to the original data. This allows the reconstruction of the original message even if some parts of the data are corrupted or lost.

The $3-bit$ repetition code is the simplest example of a classical ECC, where the encoder expands the original binary information from a single bit to three bits i.e. $0\rightarrow000$ and $1\rightarrow111$. A $3-bit$ encoder can be formalized as a mapping from the original binary, $\Theta_b$ to the logical binary codewords, $C_l$. So, when a single-bit information $'0'$ is communicated the receiver will receive $'000'$. 

\begin{equation*}
    \Theta_b = \{0,1\} \xrightarrow{\text{3-bit encoder}} C_l = \{000, 111\}
\end{equation*}

Error detection in a repetition code works by checking the bits at the receiving end. If they are not identical, the receiver knows that an error has occurred and resets the bits to the majority value. Therefore, in the case of a corrupted message, we can have three scenarios: \raisebox{.5pt}{\textcircled{\raisebox{-.9pt} {1}}} \textit{Single-bit flip error:} the receiver receives '010' instead of '000' considering that the second bit was flipped. In this case, the original codeword is generated using majority vote i.e. the corrupted message has '0' in two out of three bits, thus using majority vote the original message must be '000'. \raisebox{.5pt}{\textcircled{\raisebox{-.9pt} {2}}} \textit{Two-bit flip error:} the receiver receives '011' instead of '000' considering that the last two bits were flipped. The majority distance in this case will lead to the wrong result. \raisebox{.5pt}{\textcircled{\raisebox{-.9pt} {3}}} \textit{Three-bit flip error:} the receiver receives '111' instead of '000' considering all the bits were flipped. In this case, the receiver will not even be able to detect the error.

The distance of a code is the smallest number of bits needed to transform one codeword to another. Formally, hamming distance \cite{hamming1950error} (or distance) between two codewords $C_i$ and $C_j$ is defined as $\delta(C_i, C_j) = 2t + 1$, where $C_i, C_j \in C_l$ and $t$ is the maximum number of errors the code can correct. The maximum number of errors that can be detected by a repetition code is $\delta - 1$ and the maximum number of errors that can be corrected is $\lfloor (\delta-1)/2 \rfloor$, since the majority vote reset scheme will not work beyond this point. Therefore, for a $3-bit$ repetition code, $\delta = 3$. Traditionally, an ECC is described using the notation: $[n,k,\delta]$ where $n$ is the number of bits in a codeword, $k$ is the number of encoded bits or the original bitstring length, and $\delta$ is the code distance. For a $3-bit$ repetition code, the number of bits in the codeword is $n = 3$, the number of encoded bits is $k = 1$, and the distance is $\delta = 3$ as it requires a maximum of 3-bit flips to transform $'000'$ to $'111'$ and vice-versa. Therefore, the $3-bit$ repetition code is labeled as the $[3,1,3]$. In general, a classical $n-bit$ repetition code is labeled as $[n,1,n]$.

\paragraph{\textbf{Footsteps to Quantum Error Correction}} There exist several reasons why the direct translation of classical Error Correction Codes (ECC) into the quantum domain is non-trivial. Firstly, quantum states cannot be duplicated similarly to classical information due to the no-cloning theorem \cite{wootters1982single}. 
Secondly, qubits are vulnerable to bit-flip errors as well as phase-flip errors as mentioned in subsection \ref{qc_errors}, unlike the classical domain where bit-flip errors are the only kind of errors. Therefore, a QEC code (QECC) should be able to both detect and correct phase-flip errors along with bit-flip errors. Finally, every time a qubit is measured, the wavefunction collapses \cite{von2018mathematical} and the qubit loses its original state. Therefore, measuring a qubit directly is also not an option. An optimal QECC should possess the capability to identify and rectify both bit-flip and phase-flip errors while circumventing the direct duplication of the initial quantum state or the direct measurement of the qubits.

Classical $3-bit$ repetition code works by encoding a single bit into three bits i.e. $0\rightarrow000$ and $1\rightarrow111$. To explain QEC, we discuss the $3-qubit$ quantum repetition code first, which serves as the quantum counterpart to the classical $3-bit$ repetition code. The general idea of an $n-qubit$ repetition code is that the original state $\ket{\psi}$ is encoded with $n$ qubits to form a logical state, $\ket{\psi}_l$ \cite{gottesman1997stabilizer}. This distributes the quantum information, $\ket{\psi}$, across the entangled logical state, $\ket{\psi}_l$. Formally a $3-qubit$ quantum code is represented as,
\begin{equation*}
    \begin{split}
        \ket{\psi}_\text{ } & = \alpha\ket{0} + \beta\ket{1} \xrightarrow{\text{3-qubit encoder}} \ket{\psi}_l \\ 
        \ket{\psi}_l & = \alpha\ket{0}_l + \beta\ket{1}_l \\
        & = \alpha\ket{0} \otimes \ket{0} \otimes \ket{0} + \beta\ket{1} \otimes \ket{1} \otimes \ket{1} \\
        & = \alpha\ket{000} + \beta\ket{111} \\
        & \text{\textit{where:} } \ket{0}_l = \ket{000}; \ket{1}_l = \ket{111}
    \end{split}    
\end{equation*}

\begin{figure}
    \centering
    \includegraphics[width=0.8\linewidth]{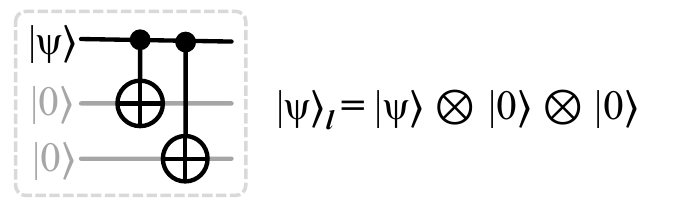}
    \caption{Illustration of the quantum circuit utilized to create $\ket{\psi}_l$ from the original state $\ket{\psi}$ in a $3-qubit$ repetition code.}
    \label{fig:qec_init}
\end{figure}

Due to the prohibition imposed by the no-cloning theorem \cite{wootters1982single}, 
the encoding of the state $\ket{\psi}$ is done by applying CNOT gates to prepare the logical state, $\ket{\psi}_l$. Fig. \ref{fig:qec_init} shows the quantum circuit that is used to expand the original state $\ket{\psi}$ into its logical state, $\ket{\psi}_l$.

Similar to classical ECC, QEC codes are implemented in a three-step process: error detection, error deduction, and error correction. Therefore, every QEC circuit must contain all of these three components. After preparing the logical state $\ket{\psi}_l$, we can move on to error detection, which in most QEC schemes, is done using the stabilizer codes. 
In simple terms, stabilizer codes check the parity of two or more qubits using CNOT gates. The output of the stabilizers is called the syndrome bits which is $+1$ ($-1$) for an error-free (erroneous) case. Error detection is done based on the syndrome bits as it provides the location of the error. Once we know the error location we can simply correct the error. Since the Pauli gates are self-inverse \cite{witten1989quantum} applying a Pauli-gate twice returns the original state. Therefore, error correction, once we know where the error has occurred is easy - we simply have to re-apply the gate on the affected qubit. Let's visualize this with an example: let there be a single bit-flip error, $e$ on the encoded state, $\ket{\psi}_l$, such that $e = X \otimes I \otimes I$, and let there be a correction operator, $c$, such that $ce\ket{\psi}_l = \ket{\psi}_l$. Given that the Pauli-gates are self-inverse, this is satisfied when $c = e$. The following equations demonstrate the manner in which the correction operator transforms the erroneous state back to its accurate state.

\begin{equation*}
\begin{split}
    &\ket{\psi}_l \xrightarrow{\text{\textit{error}}} e\ket{\psi}_l = (X \otimes I \otimes I)\ket{\psi}_l = \alpha\ket{100} + \beta\ket{011} \\
    &e\ket{\psi}_l \xrightarrow{\text{\textit{correction}}} ec\ket{\psi}_l = (X \otimes I \otimes I)(X \otimes I \otimes I)\ket{\psi}_l \\ 
    &= \alpha\ket{000} + \beta\ket{111} = \ket{\psi}_l
\end{split}
\end{equation*}

Say, we want to make parity measurements on the encoded state, $\ket{\psi}_l = \alpha\ket{000} + \beta\ket{111}$, there are three possible parities : ($Z \otimes Z \otimes I$), ($Z \otimes I \otimes Z$) and ($I \otimes Z \otimes Z$). All three of these will result in $+1$. Now suppose, we have a bit flip error on the second qubit, so $e\ket{\psi}_l = \alpha\ket{010} + \beta\ket{101}$, two out of three parity checkers measurement will return $-1$. Different errors on qubits respond to different combinations of pairwise parity measurement. Table \ref{tab:parity_op} \cite{danbrowne2021ecc} shows the response of various errors on a 3-qubit system with all possible parity measurement operators. Two operators $o_i, o_j$ are said to be commuting if $[o_i,o_j] = 0$,  i.e. $(o_i \otimes o_j) = (o_j \otimes o_i)$ and they are anti-commuting if $[o_i,o_j] \neq 0$,  i.e. $(o_i \otimes o_j) = - (o_j \otimes o_i$). Therefore, if the outcome of Table \ref{tab:parity_op} is $+1$, the error commutes with the parity operator and if the outcome is $-1$, the error anti-commutes with the parity operator. We will later use this concept in stabilizer formalism.

\begin{figure}[t]
    \centering
    \includegraphics[width=0.9\linewidth]{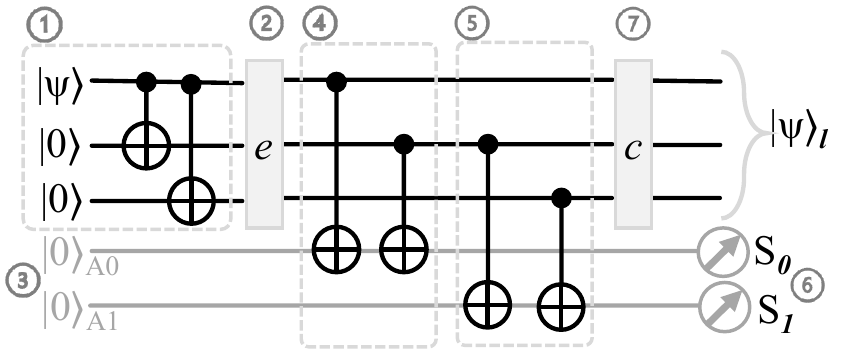}
    \caption{Demonstrating the quantum circuit that implements a $3-qubit$ repetition code. The circuit comprises the following components:
    \raisebox{.5pt}{\textcircled{\raisebox{-.9pt} {\textbf{1}}}}: The state preparation circuit represented in  Fig. \ref{fig:qec_init}.
    \raisebox{.5pt}{\textcircled{\raisebox{-.9pt} {\textbf{2}}}}: A single bit-flip error that may occur on any of the 3 qubits. 
    \raisebox{.5pt}{\textcircled{\raisebox{-.9pt} {\textbf{3}}}}: Two ancilla qubits that are initialized to the state, $\ket{0}$ are employed for parity checking purposes.
    \raisebox{.5pt}{\textcircled{\raisebox{-.9pt} {\textbf{4}}}}: The first stabilizer circuit is responsible for measuring the parity between the first two qubits.
    \raisebox{.5pt}{\textcircled{\raisebox{-.9pt} {\textbf{5}}}}: The second stabilizer circuit measures the parity between the last two qubits.
    \raisebox{.5pt}{\textcircled{\raisebox{-.9pt} {\textbf{6}}}}: The ancilla qubits are utilized to obtain the syndrome bits, $S_0, S_1$, which enable the detection and deduction of errors. 
    \raisebox{.5pt}{\textcircled{\raisebox{-.9pt} {\textbf{7}}}}: To rectify the errors, a correction operator, $c$, comprising of a sequence of self-inverse Pauli-gates, is applied to the qubit that requires correction.
    }
    \label{fig:3qec_rep_code}
    \vspace{-5mm}
\end{figure}

\begin{table}[t]
\caption{Eigenvalues of various errors on a 3-qubit system with respect to parity measurement operators\cite{danbrowne2021ecc}.}
\begin{center}
\begin{tabular}{||c || c | c | c||} 
 \hline
 Error & $Z \otimes Z \otimes I$ & $Z \otimes I \otimes Z$ & $I \otimes Z \otimes Z$ \\ [0.5ex] 
 \hline\hline
 $I \otimes I \otimes I$ & $+1$ & $+1$ & $+1$ \\
 \hline
 $X \otimes I \otimes I$ & $-1$ & $-1$ & $+1$ \\ 
 \hline
 $I \otimes X \otimes I$ & $-1$ & $+1$ & $-1$ \\
 \hline
 $I \otimes I \otimes X$ & $+1$ & $-1$ & $-1$ \\
 \hline
\end{tabular}
\label{tab:parity_op}
\vspace{-7mm}
\end{center}
\end{table}

Fig. \ref{fig:3qec_rep_code} shows the entire circuit of a $3-qubit$ repetition code which is divided into seven parts. \raisebox{.5pt}{\textcircled{\raisebox{-.9pt} {\textbf{1}}}}: First, state preparation is done which produces the logical state $\ket{\psi}_l = \ket{000}$ from the original state, $\ket{\psi} = \ket{0}$. {\textcircled{\raisebox{-.9pt} {\textbf{2}}}}: A single qubit bit-flip error $e$ occurs on one of the three qubits in the logical state, $\ket{\psi}_l$ and produces an erroneous state, $e\ket{\psi}_l$. {\textcircled{\raisebox{-.99pt} {\textbf{3}}}}: Two ancilla qubits $A_0,A_1$ are initialized to state $\ket{0}$. {\textcircled{\raisebox{-.9pt} {\textbf{4}}}}: The first stabilizer measures $Z \otimes Z \otimes I$, which is equivalent to the parity between the first two qubits by using CNOT gates among the qubits and the first ancilla qubit. {\textcircled{\raisebox{-.99pt} {\textbf{5}}}}: The second stabilizer measures $I \otimes Z \otimes Z$, which is equivalent to the parity between the last two qubits by using CNOT gates among the qubits and the second ancilla qubit. {\textcircled{\raisebox{-.9pt} {\textbf{6}}}}: The syndrome bits $S_0, S_1$ measure the pauli-z expectation values of the ancilla qubits $A_0, A_1$ respectively. The value of $S_i$ is either $+1$ or $-1$ depending on whether or not any error has occurred. Table \ref{tab:qec_3qbit_syn} shows the value of the syndrome bits with respect to the erroneous state, $e\ket{\psi}_l$. {\textcircled{\raisebox{-.9pt} {\textbf{7}}}}: Once we know where and which error has occurred, we simply apply the same gate on the affected qubit as the Pauli-gates are self-inverse. Table \ref{tab:qec_3qbit_syn} also shows how the Pauli-gates can be applied to correct the errors in the state $e\ket{\psi}_l$, for example, if the erroneous state is $\ket{100}$, then the correction operator is $XII$. Since the stabilizers are used to measure parity and the actual measurements are done on the ancilla qubits, the logical state $\ket{\psi}_l$ remains unharmed. Consequently, we have obtained a code capable of detecting and rectifying X-flip errors without adversely affecting the state of the system.

\begin{table}[t]
\caption{Detection, Deduction, and Correction of errors with respect to the syndrome measurements.}
\begin{center}
\begin{tabular}{||c | c | c | c | c||} 
\hline
 \multicolumn{2}{|c|}{\textbf{Detection}} & \multicolumn{2}{|c|}{\textbf{Deduction}} & \multicolumn{1}{|c|}{\textbf{Correction}}\\
 \hline
 \hline
 $S_0$ & $S_1$ & Error Location & Erroneous State & Correction Operator \\ [0.5ex] 
 \hline\hline
 $+1$ & $+1$ & No error & $\ket{000}$ & $III$ \\ 
 \hline
 $-1$ & $+1$ & Qubit 1 & $\ket{100}$ & $XII$ \\
 \hline
 $-1$ & $-1$ & Qubit 2 & $\ket{010}$  & $IXI$ \\
 \hline
 $+1$ & $-1$ & Qubit 3 & $\ket{001}$ & $IIX$ \\
 \hline
\end{tabular}
\label{tab:qec_3qbit_syn}
\vspace{-7mm}
\end{center}
\end{table}


In a manner analogous to single-qubit operators, such as $X$, $Z$, and others, which act on the state $\ket{\psi}$, there exist logical operators that act on the logically encoded state $\ket{\psi}_l$. These operators, known as the $logical-X$ operator ($\Bar{X}$) and $logical-Z$ operator ($\Bar{Z}$), execute bit-flip or phase-flip operations on the entire encoded state, rather than solely on individual physical qubits. By doing so, they preserve the error-correcting properties of the encoded system while facilitating the execution of logical operations as needed. The distance of a quantum code is the size of the logical Pauli operator that can transform one codeword into another. Therefore, by intuition the logical Pauli-X operator should be $\Bar{X} = X \otimes X \otimes X$ i.e. $\Bar{X}\ket{000} = \ket{111}; \Bar{X}\ket{111} = \ket{000}$. If the quantum circuits is only been susceptible to Pauli-X errors, the distance of the 3-qubit repetition code would be 3. However, we also have to consider the logical Pauli-Z operator. Following the footsteps of the logical Pauli-X operator, we can write the logical Pauli-Z operator as $\Bar{Z} = Z \otimes Z \otimes Z$ i.e. $\Bar{Z}\ket{000} = \ket{000}; \Bar{Z}\ket{111} = -\ket{111}$. However, the same can be achieved only by using one Z-operator i.e. by using $\Bar{Z} = Z \otimes I \otimes I$. The proof below shows that the distance of a 3-qubit repletion code is $1$, owing to the fact that the minimum number of operators required to transform one codeword to another is $1$. Note that the two codewords of a $3-qubit$ QECC are: $\alpha\ket{000} \pm \beta\ket{111}$.

\begin{equation*}
\vspace{-1mm}
    \begin{split}
        \Bar{Z}\ket{\psi}_l & = (Z \otimes I \otimes I)(\alpha\ket{000} + \beta\ket{111}) \\
        & = \alpha(\ket{0} \ket{0} \ket{0}) + \beta(-\ket{1} \ket{1} \ket{1}) \\
        & = \alpha\ket{000} - \beta\ket{111}
    \end{split}
\end{equation*}

A QECC is labeled as $[[n,k,\delta]]$, where $n$ is the total number of qubits, $k$ is the original number of qubits and $\delta$ is the quantum code distance. Therefore, a $3-qubit$ repetition code is labelled as $[[3,1,1]]$. The 3-qubit repetition code does not constitute a comprehensive QEC solution, as it neither detects phase flip errors nor bit-flip errors occurring on multiple qubits. This code primarily serves as a means to emphasize the essential components necessary for constructing robust and comprehensive error correction codes.

\section{Fundamentals of QEC} \label{qec_fundamentals}

Most QEC circuits work in 5 steps: a state preparation, a stabilizer circuit, error detection, an error decoder, and finally error correction. So far we have seen the basics of all of these five steps using the $3-qubit$ bit-flip error correcting code shown in Fig. \ref{fig:3qec_rep_code}. \raisebox{.5pt}{\textcircled{\raisebox{-.9pt} {1}}} in the figure shows the state preparation where a given basis state is entangled with two arbitrary qubits. \raisebox{.5pt}{\textcircled{\raisebox{-.9pt} {4}}} and \raisebox{.5pt}{\textcircled{\raisebox{-.9pt} {5}}} show the sets of stabilizers that can be used in a $3-qubit$ ECC. Based on the output of the stabilizer circuits we perform: error detection, error deduction, and error correction. Error is detected if either of the syndromes measures $-1$. Based on the values of the syndrome, the error is deducted i.e. where the error has occurred and based on this deduction the correction operator, $c$ is applied. This completes a full ECC. In this section, we concentrate on the fundamental building blocks required for the development of a foolproof QECC.

\subsection{Stabilizer Formalism}

Most QECC use stabilizers \cite{gottesman1997stabilizer} to perform error detection. Therefore, a generalized idea of forming stabilizers irrespective of the code should be useful. 
A stabilizer is defined to be a set of operators that leave a quantum state unchanged i.e. $o\ket{\psi} = \psi$ where $o$ is a set of stabilizer operators and $\ket{\psi}$ is a quantum state. Stabilizers are an important part of ECC primarily because it detects the error without harming or changing the original basis state.

Stabilizer formalism is an efficient way to manipulate and describe quantum states when it comes to the context of QECCs and fault-tolerant quantum computing. It is a mathematical framework that is generated by tensor products of the Pauli matrices (shown below):
\begin{equation*}
    \textbf{I = }
    \begin{bmatrix}
        1 & 0\\
        0 & 1
    \end{bmatrix};
    \textbf{X = }
    \begin{bmatrix}
        0 & 1\\
        1 & 0
    \end{bmatrix};
    \textbf{Y = }
    \begin{bmatrix}
        0 & -i\\
        i & 0
    \end{bmatrix};
    \textbf{Z = }
    \begin{bmatrix}
        1 & 0\\
        0 & -1
    \end{bmatrix}    
\end{equation*}

The Pauli group is a group of Pauli matrices that act on $n$ qubits and is generated by the tensor products of the above-mentioned matrices. It has $2*4^n$ elements, which include a global phase factor of $\pm 1$ or $\pm i$. For example, the Pauli group, $P$ that acts on a single qubit will contain the following elements: $P = \{ \pm I, \pm iI,  \pm X, \pm iX,  \pm Y, \pm iY,  \pm Z, \pm iZ \}$.

A stabilizer generator (or stabilizer) that acts on $n$ qubits is a product of a maximal set of $n$ commuting elements of the Pauli group, $P$ with an eigenvalue of $+1$. These $n$ operators are denoted as $g_1, g_2, \dots, g_n$. Therefore, an element, $g$, in an abelian group of self-inverse Pauli operators should satisfy the following relation: $g = \Pi g_i \ket{\psi} = \ket{\psi}; i = 1,2, \dots, n$.

Stabilizer generators must obey following relations:

\begin{itemize}
    \item Each generator is an element of the Pauli group.
    \item They commute with each other i.e. $g_i \otimes g_j = g_j \otimes g_i; \forall i,j$
    \item All generators are independent of each other meaning, no product of a subset of the generators can be equal to another generator. In simpler terms, we cannot create one by multiplying others.
\end{itemize}

\begin{figure}[t]
    \centering
    \includegraphics[width=0.9\linewidth]{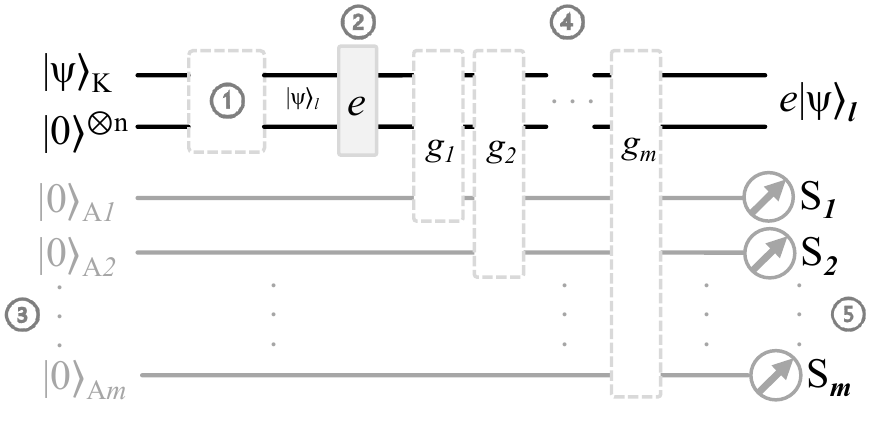}
    \vspace{-4mm}
    \caption{Illustration of a generalized stabilizer circuit that encodes $k$ physical qubits with $n$ logical qubits and employs $m$ stabilizers to detect potential errors in the logical codeword, $\ket{\psi})_l$, where $m=n-k$. The circuit comprises the following components:
    \raisebox{.5pt}{\textcircled{\raisebox{-.9pt} {\textbf{1}}}}: The encoder circuit that encodes $k$ physical qubits with $n$ logical qubits.
    \raisebox{.5pt}{\textcircled{\raisebox{-.9pt} {\textbf{2}}}}: A bit-flip or phase-flip or both errors that may occur on one or more qubits within the logical codeword, $\ket{\psi}_l$.
    \raisebox{.5pt}{\textcircled{\raisebox{-.9pt} {\textbf{3}}}}: $m$ ancilla qubits that are initialized to $\ket{0}$ to facilitate the projection of the measurements of the stabilizer generators.
    \raisebox{.5pt}{\textcircled{\raisebox{-.9pt} {\textbf{4}}}}: $m$ stabilizer generators, designated as as $g_1, g_2, \dots, g_m$, are employed.
    \raisebox{.5pt}{\textcircled{\raisebox{-.9pt} {\textbf{5}}}}: Each stabilizer $g_i$ projects its measurements onto the ancilla qubit $\ket{0}_{Ai}$. The ancilla qubits are subsequently measured as syndrome measurements, $S_i$. Based on these syndrome values, error deduction and correction mechanisms are executed. It is essential to highlight that the provided diagram primarily illustrates a generalized stabilizer circuit designed for detecting errors. This representation does not encompass the execution of correction operators; consequently, the final basis state persists as the erroneous state, $e\ket{\psi}_l$.
    }
    \vspace{-5mm}
    \label{fig:stab_form}
\end{figure}

With a basic mathematical background on Pauli operators and stabilizer generators, we can define a stabilizer code which is a quantum error-detecting code that is formed by a set of stabilizer generators. This code encodes $k$ qubits into $n$ physical qubits. For each stabilizer code, there are encoded logical operators that depict operations on the encoded qubits. These logical operators are denoted as $\Bar{X_i}$ and $\Bar{Z_i}$ for $i = 1,2, \dots, k$. It is important to note that these logical operators do not change the stabilizer generators, they only have an effect on the encoded qubits and they follow the same rules as that of the standard $X$ and $Z$ operators.

Theoretically, every stabilizer is dividing the Hilbert space based on the eigenvalue ($\pm1$). The size of the Hilbert space at the beginning is $2^n$ as we are using $n$ physical qubits. If we apply $m$ stabilizer generators, the Hilbert space is getting divided and finally, it boils down to the size of $\frac{2^n}{2^m}$. We already know that we are encoding $n$ physical qubits with $k$ logical qubits, therefore: $\frac{2^n}{2^m} = 2^k \Rightarrow m = n-k$.

From the above equation it is important to note that if one increases the number of stabilizer generators, the number of logical qubits may decrease. Let us now see what a generalized stabilizer circuit looks like: there is basis state, $\ket{\psi}_K$ with $k$ qubits i.e. $\ket{\psi}_K = \ket{\psi_1\psi_2\dots\psi_k}$ that will be encoded with $n$ logical qubits, $\ket{0}^{\otimes n} = \ket{000\dots}$ and $m$ stabilizers will be applied on the final logical codeword, $\ket{\psi}_l$. Fig. \ref{fig:stab_form} shows a generalized stabilizer circuit that detects errors.

\begin{figure}[t]
    \centering
    \includegraphics[width=1\linewidth]{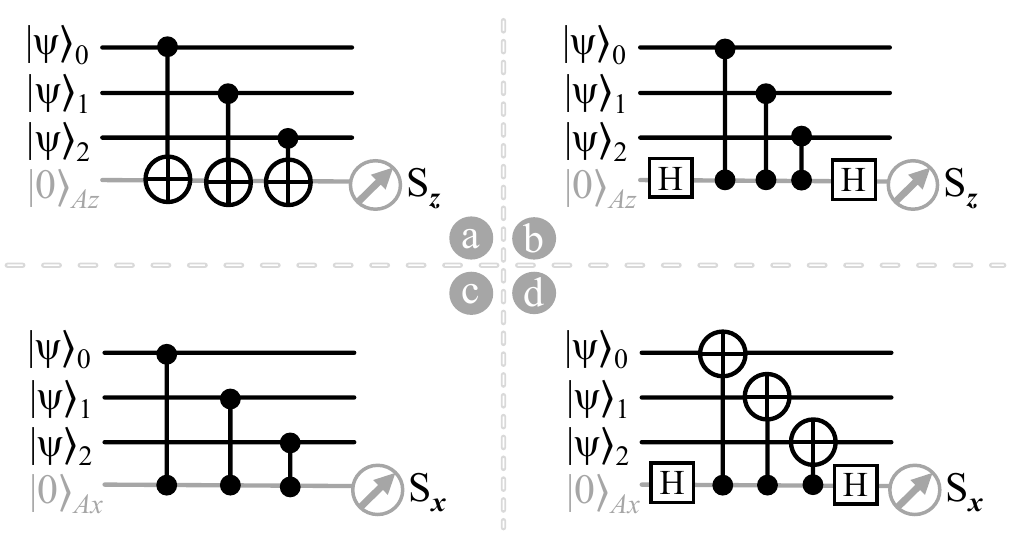}
    \vspace{-8mm}
    \caption{Different approaches to generate a $Z-stabilizer$ and $X-stabilizer$ operating on 3 qubits, namely $\ket{\psi}_0$, $\ket{\psi}_1$ and $\ket{\psi}_2$. The ancilla qubit, on which the stabilizer output is being projected, is represented as $\ket{0}_{Az}$ and $\ket{0}_{Ax}$ for $Z-stabilizers$ and $X-stabilizers$, respectively. The circuits are presented as follows:
    \raisebox{.5pt}{\textcircled{\raisebox{-.2pt} {\textbf{a}}}}:  This circuit exhibits a $Z-stabilizer$ similar to the one depicted in Fig. \ref{fig:3qec_rep_code}. The ancilla qubit, $\ket{0}_{Az}$ measures $Z\ket{\psi}_0 \otimes Z\ket{\psi}_1 \otimes Z\ket{\psi}_2$, commonly denoted as $Z_0 \otimes Z_1 \otimes Z_2$.
    \raisebox{.5pt}{\textcircled{\raisebox{-.9pt} {\textbf{b}}}}: This circuit demonstrates the same $Z-stabilizer$ as in (a). The stabilizer is regenerated using $H$ gates and $CZ$ gates instead of $CNOT$ gates. The ancilla qubit, $\ket{0}_{Az}$ also measures $Z_0 \otimes Z_1 \otimes Z_2$.
    \raisebox{.5pt}{\textcircled{\raisebox{-.2pt} {\textbf{c}}}}: This circuit represents an $X-stabilizer$. The ancilla qubit, $\ket{0}_{Ax}$ measures $X\ket{\psi}_0 \otimes X\ket{\psi}_1 \otimes X\ket{\psi}_2$, commonly denoted as $X_0 \otimes X_1 \otimes X_2$.
    \raisebox{.5pt}{\textcircled{\raisebox{-.9pt} {\textbf{d}}}}: This circuit portrays the same $X-stabilizer$ as mentioned in (c). he stabilizer is recreated using $H$ gates and $CNOT$ gates instead of $CZ$ gates. The ancilla qubit, $\ket{0}_{Ax}$ also measures $X_0 \otimes X_1 \otimes X_2$.
    }
    \label{fig:all_stab}
    \vspace{-4.5mm}
\end{figure}

\subsection{Structure of the Stabilizer Generator}

\begin{figure}[t]
    \centering
    \includegraphics[width=1\linewidth]{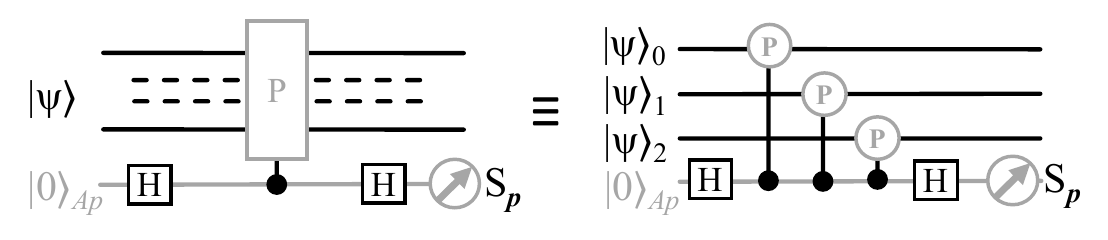}
    \vspace{-8mm}
    \caption{Illustration of a universal circuit of a stabilizer generator, in which $P$ represents a Pauli-gate operation. This circuit measures $P\ket{\psi}_0 \otimes P\ket{\psi}_1 \otimes P\ket{\psi}_2 \equiv P_0 \otimes P_1 \otimes P_2$. Usually for a stabilizer generator, $P \in \{X, Z\}$.}
    \label{fig:stab_gen}
    \vspace{-5mm}
\end{figure}

For a stabilizer generator (or stabilizer) to be able to detect an error, the stabilizer has to anti-commute with the error (see Table \ref{tab:parity_op}). Therefore, we need $Z-stabilizers$ to detect the $X$ or $bit-flip$ errors and $X-stabilizers$ to detect $Z$ or $phase-flip$ errors. A $Z-stabilizer$ will anti-commute with a $X$ or $bit-flip$ error and produce an eigenvalue of $-1$.
In Fig. \ref{fig:3qec_rep_code}: \raisebox{.5pt}{\textcircled{\raisebox{-.9pt} {4}}} and \raisebox{.5pt}{\textcircled{\raisebox{-.9pt} {5}}} depict two $Z-stabilizers$ which can detect $X$ errors only. Let's take a quantum system with three qubits $\ket{\psi}_0$, $\ket{\psi}_1$ and $\ket{\psi}_2$, with a $Z-stabilizer$ and an $X-stabilizer$ acting on these three qubits. Finally, these stabilizers will project their values onto the ancilla qubits $\ket{0}_{Az}$ and $\ket{0}_{Ax}$, respectively. In Fig. \ref{fig:all_stab} we note that $\ket{0}_{Az}$ measures $Z_0 \otimes Z_1 \otimes Z_2$ which will anti-commute with $X-errors$ to produce an eigenvalue of $-1$. Similarly, $\ket{0}_{Ax}$ measures $X_0 \otimes X_1 \otimes X_2$ which will anti-commute with $Z-errors$ to produce an eigenvalue of $-1$. Fig. \ref{fig:all_stab} also shows different ways of creating the two different types of stabilizers \cite{google2023suppressing}. A generalized circuit of a stabilizer generator is shown in Fig. \ref{fig:stab_gen} where $P$ is a Pauli-gate operation and the circuit measures $P\ket{\psi}_0 \otimes P\ket{\psi}_1 \otimes P\ket{\psi}_2 \equiv P_0 \otimes P_1 \otimes P_2$. An ancilla qubit, initialized to $\ket{0}$ is a control for an arbitrary state $\ket{\psi},$ using a unitary operator, $P$ i.e., operator $P$ is applied when the control is $\ket{1}$ and nothing happens when the control is $\ket{0}$. We explain the state of the circuit (Fig. \ref{fig:stab_gen}: \textit{left}), one step at a time:
\vspace{-6mm}

\begin{equation*}
    \begin{split}
        \text{\textit{First H gate }} & \Rightarrow \frac{1}{\sqrt{2}}\bigl(\ket{0}+\ket{1}\bigr)\ket{\psi} \\
        \text{\textit{Controlled P }} & \Rightarrow \frac{1}{\sqrt{2}} \bigl(\ket{0}\ket{\psi} + \ket{1}P\ket{\psi}\bigr) \\
        \text{\textit{Second H gate }} & \Rightarrow \frac{1}{2} \Bigl(\bigl(\ket{0}+\ket{1}\bigr)\ket{\psi} + \bigl(\ket{0}-\ket{1}\bigr)P\ket{\psi}\Bigr) \\
        & = \ket{0}\biggl(\frac{1+P}{2}\biggr)\ket{\psi} + \ket{1}\biggl(\frac{1-P}{2}\biggr)\ket{\psi} \\ 
        \text{\textit{Finally, }}\ket{0}_{Ap} & = \ket{0}\lambda_{+}^{P}\ket{\psi} + \ket{1}\lambda_{-}^{P}\ket{\psi} \\
        \text{\textit{where the }} & \text{\textit{eigenstate projectors are }} \Rightarrow \lambda_{+}^{P}; \lambda_{-}^{P}
    \end{split}
\end{equation*}

The primary importance of the above equations is to show that the stabilized state, $\ket{\psi}$ will always hold the form $\alpha '\ket{0} + \beta '\ket{1}$. Consequently, even if an error occurs at an arbitrary rotational angle, the stabilizer circuit will enforce a non-superposed state for the error. As a result, when the error detection circuit is applied, the stabilizers will consistently return an eigenstate of either $\pm 1$ \cite{nielsen2002quantum}.

\subsection{Topological Codes}

Repetition codes exhibit limitations as QECCs because of their restricted error correction capacities and diminished error thresholds. Moreover, they do not possess fault tolerance and encounter scalability issues. These codes are also unsuitable for rectifying errors that transpire on multiple qubits or complex errors. Therefore, a class of QEC codes, known as the Topological Codes are proposed that use the properties of topological order \cite{wen2017colloquium} to protect the quantum information from errors. Topological codes \cite{kitaev1997quantum} such as, toric codes are designed to process and store quantum information by exploiting the global features of lattice-like structures, which makes them naturally resistant to local errors. Topological codes have gained significant popularity due to their fault-tolerant properties and ability to correct quantum errors with relatively low overhead. Some key features of topological codes include: 

\begin{itemize}
    \item \textit{Spatial separation:} All logical qubits are encoded using non-local degrees of freedom, meaning that the probability of local errors corrupting the information is really low. Simply put, the logical qubits are spread out (i.e., not close to each other) making it harder for local errors to affect the information encoded.
    \item \textit{Anyonic excitation:} So far topological codes are built using anyons, which have a unique braiding property to manipulate and store quantum information
    \item \textit{Error correction:} Stabilizer checks \cite{calderbank1996good, steane1996error, tillich2013quantum, kovalev2013quantum} are an important part of error correcting codes but stabilizers in topological codes are non-local operators and they can detect the anyonic excitation created by errors.
    \item \textit{Fault tolerant quantum computation:} Topological codes support fault-tolerant quantum computing with a low error rate even in the presence of gate errors.
\end{itemize}

There exist multiple types of Topological codes, which can include, but are not limited to, the following:

\begin{itemize}
    \item \textit{Toric codes:} These are one of the earliest known topological codes and are defined in a two-dimensional lattice with periodic boundary conditions. The robustness of detecting and correcting errors is increased by the structure of the lattice \cite{kitaev2003fault}.
    \item \textit{Surface codes:} These are also one of the earliest known topological codes and are defined in a two-dimensional lattice, but unlike the toric codes, without periodic boundary conditions. They are currently the most promising candidate for a large-scale, fault-tolerant quantum computer due to their high error threshold and low overhead \cite{dennis2002topological}. Realization of a surface code has been the primary goal for multiple research articles \cite{kelly2016scalable, sete2016functional, o2016silicon, krinner2022realizing, takita2017experimental}.
    \item \textit{Color codes:} These are another type of topological codes that can be either on a two or three-dimensional lattice. They possess similar error-correcting properties as surface codes but offer additional advantages like the ability to apply certain logical gates transversely \cite{bombin2006topological, bombin2007topological}.
\end{itemize}

The overarching structure of a topological code entails the construction of the complete code by assembling repetitive elements. In this subsection, we primarily discuss the functioning of topological codes, with a particular emphasis on toric and surface codes. The operation of color codes and other topological codes is beyond the scope of this paper.

\begin{figure}
    \centering
    \includegraphics[width=1\linewidth]{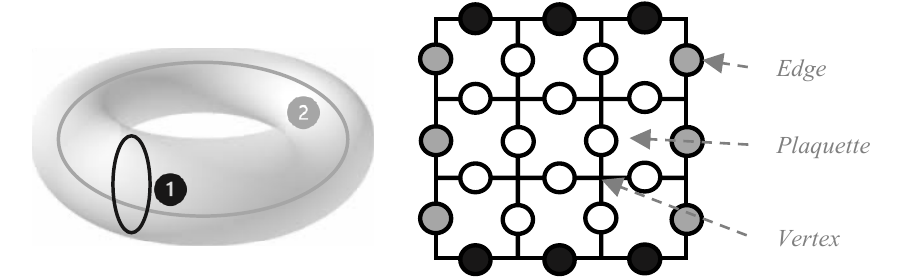}
    \vspace{-6mm}
    \caption{\textbf{\textit{Left:}} A torus that is used to model a toric code. \textbf{\textit{Right:}} A toric code in the form of a lattice. The lattice comprises of two loops that run through the torus. Various components of the lattice, such as the edge, plaquette, and vertex, are designated in the figure.
    \raisebox{.5pt}{\textcircled{\raisebox{-.9pt} {\textbf{1}}}} represents the vertical loop that corresponds to the top-most and the lower-most edge of the lattice, whereas \raisebox{.5pt}{\textcircled{\raisebox{-.9pt} {\textbf{2}}}} represents the horizontal loop that corresponds to the left-most and the right-most edge of the lattice. It is noteworthy that a surface code would possess a lattice that appears exactly like the one depicted here but would not have any boundary conditions. This means that a surface code does not model on a torus and is therefore referred to as a planar code.
    }
    \vspace{-6mm}
    \label{fig:toric_basic}
\end{figure}

The toric code is defined in terms of a square lattice with periodic boundary conditions, meaning if we have an $L^*L$ square lattice wrapped around a torus, then the right-most edge is equivalent to the left-most edge and the top-most edge is equivalent to the lower-most edge. For surface codes such boundary conditions do not exist and hence it is often referred to as the planar code \cite{bravyi1998quantum, freedman2001projective}. Fig. \ref{fig:toric_basic} depicts a torus and how it is used to model the boundary conditions of a toric code lattice. It is important to keep in mind that the lattice for a surface code would look exactly like the toric code lattice but without the boundary conditions. A comparison between toric codes and surface codes is shown in Table \ref{tab:toric_vs_surface}.

\begin{table*}[t]
\caption{A comparative study between Toric Codes and Surface Codes}
\vspace{-4mm}
\begin{center}
\begin{tabular}{||c | c | c||} 
 \hline
 \textbf{Property} & \textbf{Toric Code} & \textbf{Surface Code}\\ [0.5ex] 
 \hline\hline
 Dimensionality & 2D in a $L^*L$ lattice & 2D in a $L^*L$ lattice \\ 
 \hline
 Lattice Structure & Regular lattice with periodic boundary conditions & Regular lattice without periodic boundary conditions\\
 \hline
 Logical Qubits & $2$ (two independent logical qubits) & $1$ (one logical qubit)\\
 \hline
 Error Correction & Detects and corrects any single-qubit error & Detects and corrects any single-qubit error\\
 \hline
 Stabilizer Generators & Two types: vertex and plaquette operators & Two types: vertex and plaquette operators\\
 \hline
 Boundary Conditions & Periodic (closed topology) & Open (open topology)\\
 \hline
 Logical Gates & Braiding anyons & Lattice surgery or code deformation\\
 \hline
 Implementation Complexity & More complex due to periodic boundary conditions & Simpler due to open boundary conditions\\
 \hline
 $[[n,k,\delta]]$ & $[[2L^2,2,L]]$ & $[[2L^2,1,L]]$\\
 \hline
\end{tabular}
\label{tab:toric_vs_surface}
\vspace{-7mm}
\end{center}
\end{table*}

\begin{figure}
    \centering
    \includegraphics[width=0.75\linewidth]{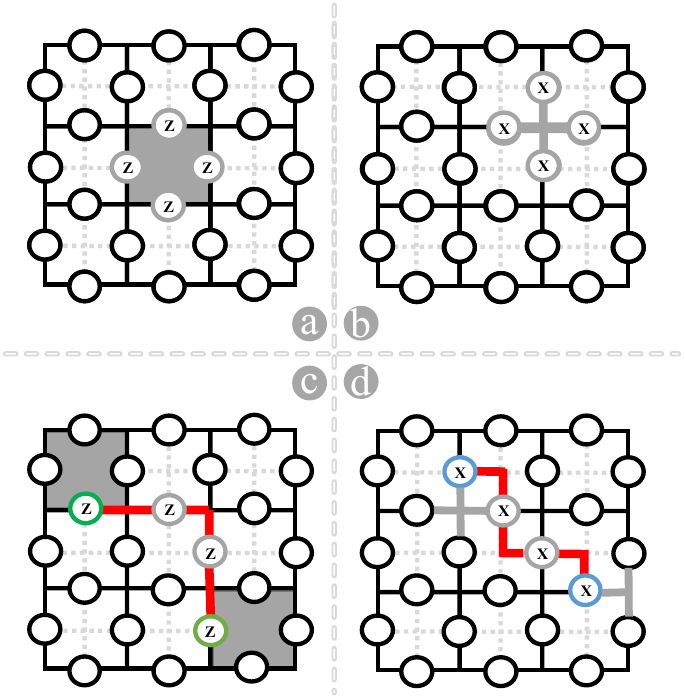}
    \vspace{-4mm}
    \caption{
    Depiction of $Z$ and $X-stabilizers$ acting on qubits in a toric code, and showcasing of the erroneous qubits that trigger plaquettes and vertices at the end of the error strings.
    \raisebox{.5pt}{\textcircled{\raisebox{-.4pt} {\textbf{a}}}}: A plaquette that represents $Z-stabilizers$ acting on qubits. \raisebox{.5pt}{\textcircled{\raisebox{-.9pt} {\textbf{b}}}}: A vertex that represents $X-stabilizers$ acting on qubits.
    {\textcircled{\raisebox{-.2pt} {\textbf{c}}}}: The red line exhibits the error string that triggers the marked $Z-operators$, However, only the $plaquette-operators$ that are marked in grey at the end of the error string will produce an eigenstate $-1$. The green color denotes the erroneous qubits that activate the grey plaquettes at the end of the error string.
    {\textcircled{\raisebox{-.9pt} {\textbf{d}}}}: The red line illustrates the error string that activates the marked $X-operators$. Nevertheless, only the $vertex-operators$ that are marked in grey at the end of the error string will generate an eigenstate of $-1$. The blue color is used to identify the erroneous qubits that initiate the grey vertices at the end of the error string.
    }
    \label{fig:lattice_stab}
    \vspace{-5mm}
\end{figure}

Every edge of a lattice corresponds to one qubit. Therefore, in a $L^*L$ lattice there are $2L^2$ edges, i.e., $2L^2$ physical qubits. Every plaquette is a $Z-stabilizer$ generator and every vertex is a $X-stabilizer$ generator. Fig. \ref{fig:lattice_stab} showcases the following aspects: \raisebox{.5pt}{\textcircled{\raisebox{-.2pt} {a}}} demonstrates that there exists a $Z-stabilizer$ that operates on the four physical qubits located on the four edges of a plaquette, while \raisebox{.5pt}{\textcircled{\raisebox{-.9pt} {b}}} depicts an  $X-stabilizer$ that operates on the four physical qubits located on the four edges of a vertex. This implies that each physical qubit would have one $Z-stabilizer$ and one $X-stabilizer$ operating on it. Hence, it is essential to note that plaquettes and vertices overlap with each other, and as a result, these two operators commute. Upon multiplying a set of plaquette operators, the resultant operator would be equal to the $Z-operators$ that operate on the boundary of the consolidated plaquettes. The same principle applies to vertices.

In the toric code, the lattice has periodic boundary conditions, meaning that it is modeled around a torus. Such a structure leads to two independent loops: one in the horizontal direction and the other in the vertical direction, as shown in Fig. \ref{fig:toric_basic}. Such a structure gives rise to two logical qubits. Mathematically, the toric code will have four logical operators: two $X-operators$ on the horizontal and vertical loops ($\bar{X1},\Bar{X2}$) and two $Z-operators$ on the horizontal and vertical loops ($\bar{Z1},\Bar{Z2}$). These operators must commute with the stabilizer generator but never with each other. Given that there are two pairs of non-commuting logical operators ($\bar{X1},\Bar{X2}$ and $\bar{Z1},\Bar{Z2}$), the toric code has $2$ logically encoded qubits. On the contrary, the surface code lattice has open boundary conditions, meaning there are no loops. In this case, there are two logical operators: one  $X-operator$ line that goes in one direction ($\Bar{X}$) and another $Z-operator$ line that goes in the direction orthogonal ($\Bar{Z}$) to the previously mentioned line. These operations must commute with the stabilizers but never with each other. Given there is only one pair of non-commuting logical operators ($\Bar{X1}$ and $\Bar{Z1}$), the surface code has only $1$ logically encoded qubit.

Every stabilizer measurement is projected onto an ancilla qubit and the syndrome measurement tells us where the error has occurred on the lattice. A stabilizer generator circuit can be found in Fig. \ref{fig:all_stab}. If in a plaquette there are four qubits, then their respective $Z-stabilizer$ or $plaquette-operator$ measurement will be $Z_1 \otimes Z_2 \otimes Z_3 \otimes Z_4$ and similarly, the $X-stabilizer$ or $vertex-operator$ will measure $X_1X_2X_3X_4$. If there is an error on any one of the qubits, depending on the type of error, either the $X-stabilizer$ or the $Z-stabilizer$ linked with that qubit will return an eigenvalue of $-1$. This is the case when only one qubit has an error.

Now suppose, there is a string of errors on the lattice, represented by a product of $Z$ operators along the string. Mathematically, when we measure the plaquette operators, we are taking the tensor product of the $Z-operators$ on the edges forming the plaquette. If the plaquette is not at the end of the string, the product of $Z-operators$ for the plaquette will cancel out and will always result in a $+1$ eigenstate. This is due to the fact that $Z-operators$ square to the identity i.e. $ZZ = I$ and the tensor product of an even number of $Z-operators$ will result in identity. However, if the plaquette is at the end of the error string, the product of $Z-operators$ will not cancel out thus, resulting in a $-1$ eigenstate. This is because there is an odd number of $Z-operators$ at the end plaquette, making the tensor products of $Z-operators$ unequal to the identity. In simple terms, when there is a string of errors on the lattice, the only plaquette or vertex with $-1$ eigenstate will be at the end of the error string. Fig. \ref{fig:lattice_stab}: \raisebox{.5pt}{\textcircled{\raisebox{-.9pt} {c}}} displays that $plaquette-operators$ situated at the end of an error chain produce an eigenvalue of  $-1$ and similarly, \raisebox{.5pt}{\textcircled{\raisebox{-.9pt} {d}}} demonstrates that $vertex-operators$ located at the end of an error chain solely produce an eigenvalue of $-1$.

Given that solely the \textit{`end of the error string'} generates a signal that an error has transpired, it is imperative to address the concern of the existence of multiple paths between the two endpoints. Therefore, a fundamental query arises, which is how to identify the precise location of the error string between two given endpoints. \textit{How do we know which is the correct path?} To solve this problem large-scale QECCs use approximate inference algorithms to determine the most likely error that might have occurred given a specific syndrome value. These methods allow for application in real-time between successive stabilizer code cycles. To date, there is no particular algorithm that does this job efficiently for all ECC. For toric and surface codes an algorithm known as the Minimum Weight Perfect Matching (MWPM) is often used for decoding \cite{edmonds1965paths, kolmogorov2009blossom}. The effectiveness of a QECC's error rate is highly dependent on the type of decoder utilized, as some approximate algorithms perform better than others. To combat this problem, error-correcting codes are simulated over multiple cycles and the syndrome is sampled to better understand a noise model. Surface codes are also simulated over multiple cycles in the hope of eventually creating fault-tolerant quantum computers \cite{krinner2022realizing,divincenzo2007effective, gidney2021stim}.

\begin{figure}[t]
    \centering
    \includegraphics[width=0.75\linewidth]{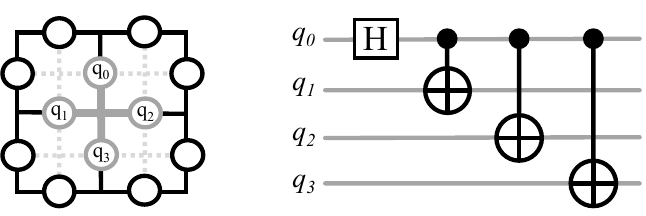}
    \vspace{-4mm}
    \caption{\textbf{\textit{Left:}} A $vertex-operator$ operator acting on four qubits, $q_0,q_1,q_2,q_3$ on a lattice. \textbf{\textit{Right:}} A state preparation encoding circuit operates on qubits $q_0,q_1,q_2,q_3$, within the vertex-operator, all initialized to $\ket{0}$. When applied to $\ket{0000}$, the circuit yields $\frac{1}{\sqrt{2}} (\ket{0000} + \ket{1111})$.}
    \label{fig:encoder}
    \vspace{-5mm}
\end{figure}

\subsection{A General Encoding Circuit}

In quantum error correction, before employing stabilizer circuits for error detection, an encoding circuit is needed to map logical qubits to physical qubits, thereby protecting quantum information. The encoding process, often termed state preparation, is represented by a unitary transformation $U$ that encodes $k$ qubits with $n$ logical qubits. Two methods exist for state preparation: stabilizer generator circuits and unitary circuits \cite{higgott2021optimal, bravyi2006lieb}. This paper focuses on the former. Encoding of logical states has been experimentally demonstrated in small-scale error correction protocols using various codes \cite{roffe2018protecting, kelly2015state, chiaverini2004realization, cramer2016repeated, gong2022experimental, linke2017fault, lu2008experimental, schindler2011experimental}.

It is well-established that a logical codeword commutes with a stabilizer generator, meaning they have $+1$ eigenstates for each stabilizer generator in a group of stabilizers. To prepare the logical state from an arbitrary input state, it is essential to project qubits into the $+1$ eigenstate of the relevant stabilizer operators. When initializing the quantum circuit with $n$ qubits, all set to $\ket{0}$, we obtain a stabilized state at $\ket{0}^{\otimes n}$. This condition is valid when applying $Z-stabilizers$; however, when applying $X-stabilizers$, nearly half of the resultant eigenstate will be $+1$, and the rest will be $-1$ due to superposition. The primary goal of the state preparation circuit is to ensure that the $X-stabilizers$ consistently return a $+1$ eigenstate in the absence of errors. Fig. \ref{fig:encoder} illustrates an $X-stabilizer$ or $vertex-stabilizer$ acting on four qubits ($q_0, q_1, q_2, q_3$) and the corresponding encoding circuit. This circuit is inspired by the Greenberger–Horne–Zeilinger (GHZ) state \cite{greenberger1989going}. Assuming the circuit depicted in Fig. \ref{fig:encoder} can be represented by U, a product of operators, and all qubits ($q_0, q_1, q_2, q_3$) initialized to $\ket{0}$, then $U\ket{0000} = \frac{1}{\sqrt{2}} (\ket{0000} + \ket{1111})$. The decomposition is valid only when the initial Hadamard qubit is in the $\ket{0}$ state; thus, the choice of Hadamard's qubit must be made cautiously. The superposition generated by the state preparation circuit counteracts the superposition induced by the $X-stabilizer$. Consequently, in the absence of errors, the X-stabilizer eigenstate yields a $+1$ result.

\section{Practicality of QECC} \label{qec_application}

In this section, we show a comparative analysis of existing QECCs in literature (Table \ref{tab:QECC_comps}). We also examine diverse qubit types and the corresponding suitable QECCs for them.  


\begin{table*}[t]
\caption{A comparative analysis among the various Quantum Error Correction Codes available in the existing literature}
\vspace{-2mm}
\begin{tabular}{||l|l|l|l|l|l|l|l||}
\hline
\multicolumn{1}{|c|}{\textbf{Code}}                                   & \multicolumn{1}{c|}{\textbf{Year}} & \multicolumn{1}{c|}{\textbf{\# Qubits}}                      & \multicolumn{1}{c|}{\textbf{Description}}                                                                                                                               & \multicolumn{1}{c|}{\textbf{Complexity}}                                                       & \multicolumn{1}{c|}{\textbf{\begin{tabular}[c]{@{}c@{}}Decoding\\ Algorithm\end{tabular}}}                   & \multicolumn{1}{c|}{\textbf{Advantage}}                                                                          & \multicolumn{1}{c|}{\textbf{Disadvantage}}                                                                        \\ \hline \hline
\begin{tabular}[c]{@{}l@{}}Shor's 9-qubit\\ code \cite{shor1995scheme}\end{tabular}         & 1995                               & 9                                                            & \begin{tabular}[c]{@{}l@{}}First quantum error-correcting\\ code, correcting 1 arbitrary\\ error. Simple example of an\\ error-correcting code.\end{tabular}            & \begin{tabular}[c]{@{}l@{}}Moderate\\ (9 physical\\ qubits for 1\\ logical qubit)\end{tabular} & \begin{tabular}[c]{@{}l@{}}Syndrome\\ measurement,\\ lookup table\end{tabular}                                & \begin{tabular}[c]{@{}l@{}}Good for\\ understanding basic\\ error correction\\ concepts.\end{tabular}            & \begin{tabular}[c]{@{}l@{}}Requires 9\\ qubits, not very\\ resource-efficient.\end{tabular}                       \\ \hline
\begin{tabular}[c]{@{}l@{}}Steane's 7-qubit\\ code \cite{steane1996simple}\end{tabular}       & 1996                               & 7                                                            & \begin{tabular}[c]{@{}l@{}}Corrects single error, example\\ of Calderbank-Shor-Steane\\ code. Exploits classical\\ error-correcting codes.\end{tabular}                 & \begin{tabular}[c]{@{}l@{}}Moderate\\ (7 physical\\ qubits for 1\\ logical qubit)\end{tabular} & \begin{tabular}[c]{@{}l@{}}Syndrome\\ measurement,\\ lookup table\end{tabular}                                & \begin{tabular}[c]{@{}l@{}}More resource-\\ efficient than Shor's\\ code, easy to\\ implement.\end{tabular}      & \begin{tabular}[c]{@{}l@{}}Only corrects\\ single error, not\\ suitable for larger\\ systems.\end{tabular}        \\ \hline
Toric codes \cite{kitaev1997quantum}                                                          & 1997                               & \begin{tabular}[c]{@{}l@{}}Varies(2D\\ lattice)\end{tabular} & \begin{tabular}[c]{@{}l@{}}Topological codes defined\\ on 2D lattice, robust\\ against local errors.\\ High threshold and\\ fault-tolerant.\end{tabular}                & \begin{tabular}[c]{@{}l@{}}High\\ (2D lattice)\end{tabular}                                    & \begin{tabular}[c]{@{}l@{}}Minimum\\ weight perfect\\ matching\\ (MWPM)\end{tabular}                          & \begin{tabular}[c]{@{}l@{}}High error\\ threshold, robust\\ against local errors.\end{tabular}                   & \begin{tabular}[c]{@{}l@{}}Requires complex\\ decoding \\ algorithms.\end{tabular}                                \\ \hline
\begin{tabular}[c]{@{}l@{}}Surface\\ codes \cite{dennis2002topological}\end{tabular}               & 2002                               & \begin{tabular}[c]{@{}l@{}}Varies(2D\\ lattice)\end{tabular} & \begin{tabular}[c]{@{}l@{}}Topological codes defined\\ on 2D lattice, most studied\\ topological code. High\\ threshold and fault-\\ tolerant.\end{tabular}             & \begin{tabular}[c]{@{}l@{}}High\\ (2D lattice)\end{tabular}                                    & \begin{tabular}[c]{@{}l@{}}Minimum\\ weight perfect\\ matching\\ (MWPM)\end{tabular}                          & \begin{tabular}[c]{@{}l@{}}High error\\ threshold, \\ well-studied, \\ many efficient\\ decoders.\end{tabular}   & \begin{tabular}[c]{@{}l@{}}Requires large\\ qubit overhead.\end{tabular}                                          \\ \hline
\begin{tabular}[c]{@{}l@{}}Bacon-Shor\\ codes \cite{bacon2006operator}\end{tabular}            & 2006                               & Varies                                                       & \begin{tabular}[c]{@{}l@{}}Subsystem codes with good\\ error-correction properties.\\ Separates error correction\\ into subsystems.\end{tabular}                        & \begin{tabular}[c]{@{}l@{}}Moderate\\ to High\\ (depends on\\ specific code)\end{tabular}      & \begin{tabular}[c]{@{}l@{}}Syndrome\\ measurement,\\ lookup table\end{tabular}                                & \begin{tabular}[c]{@{}l@{}}Simple structure,\\ transversal gates\\ for some operations.\end{tabular}             & \begin{tabular}[c]{@{}l@{}}Lower error\\ threshold\\ compared to\\ topological codes.\end{tabular}                \\ \hline
\begin{tabular}[c]{@{}l@{}}3D Color\\ codes \cite{bombin2006topological}\end{tabular}              & 2006                               & \begin{tabular}[c]{@{}l@{}}Varies(3D\\ lattice)\end{tabular} & \begin{tabular}[c]{@{}l@{}}3D generalization of surface\\ codes, improved error-\\ correction properties. \\ Combines features of toric\\ and color codes.\end{tabular} & \begin{tabular}[c]{@{}l@{}}High\\ (3D lattice)\end{tabular}                                    & \begin{tabular}[c]{@{}l@{}}Minimum\\ weight perfect\\ matching\\ (MWPM)\end{tabular}                          & \begin{tabular}[c]{@{}l@{}}Higher error\\ threshold than\\ 2D codes,\\ transversal gates.\end{tabular}           & \begin{tabular}[c]{@{}l@{}}Requires 3D\\ lattice structure,\\ more complex\\ to implement.\end{tabular}           \\ \hline
\begin{tabular}[c]{@{}l@{}}Homological\\ Product\\ codes \cite{haah2011local}\end{tabular} & 2013                               & Varies                                                       & \begin{tabular}[c]{@{}l@{}}Product codes combining\\ different quantum codes,\\ allowing transversal gates.\\ Exploits the structure of\\ different codes\end{tabular}  & \begin{tabular}[c]{@{}l@{}}Varies\\ (depends on\\ specific\\ codes used)\end{tabular}          & \begin{tabular}[c]{@{}l@{}}Syndrome\\ measurement,\\ classical error-\\ correction\\ algorithms\end{tabular}  & \begin{tabular}[c]{@{}l@{}}Enables\\ transversal gates,\\ versatile, can\\ combine various\\ codes.\end{tabular} & \begin{tabular}[c]{@{}l@{}}Complexity and\\ error threshold\\ depend on the\\ specific codes\\ used.\end{tabular} \\ \hline
\begin{tabular}[c]{@{}l@{}}Flag-qubit\\ codes \cite{chamberland2020topological}\end{tabular}            & 2020                               & Varies                                                       & \begin{tabular}[c]{@{}l@{}}This code is constructed on\\ low degree graphs.\end{tabular}                                                                                & \begin{tabular}[c]{@{}l@{}}Varies\\ (depends on\\ specific\\ parameters)\end{tabular}          & \begin{tabular}[c]{@{}l@{}}Classical\\ maximum\\ likelihood\\ decoding\end{tabular}                           & \begin{tabular}[c]{@{}l@{}}Reduces overhead\\ and maintains high\\ error threshold.\end{tabular}                 & \begin{tabular}[c]{@{}l@{}}The decoding\\ algorithm can be\\ computationally\\ demanding.\end{tabular}            \\ \hline
\end{tabular}
\label{tab:QECC_comps}
\end{table*}

\subsection{Superconducting Qubits}
\paragraph{\textbf{Technological Details}} 
Superconducting qubits are LC oscillator circuits maintained at cryogenic temperatures. Typically, the inductor is implemented as a Josephson Junction (for introducing non-linearity in the circuit) using superconducting material such as, niobium or aluminium
and the capacitor is implemented as either an inter-digitated capacitor or parallel plate capacitor. Additionally, the Josephson Junction possesses its own intrinsic capacitance. The overall LC circuit forms a harmonic oscillator that creates different energy levels out of which the lowest two energy levels are selected as the basis states of the qubit. The basis states are also determined by the flow of the current in the LC circuit e.g., flow in one direction may correspond to state $\ket{0}$ while flow in the opposite direction may correspond to state $\ket{1}$ (more information can be found in \cite{krantz2019quantum}).

\paragraph{\textbf{Feasibility}}
Surface codes are implemented in \cite{andersen2020repeated} on seven qubits (four data qubits, three ancilla qubits). It has been observed that repeated error correction results in longer coherence times of qubits than no error correction at a high 96.1\% average local fidelity. \cite{zhao2022realization} implements a distance three surface code on 17 qubits to achieve fidelity up to 0.9 after error correction. A [[5,1,3]] error correcting code is produced in \cite{gong2019experimental} using 92 gates overall that corrects single-qubit gate errors at $\sim$75\% fidelity. There are older works like \cite{kelly2015state} that incorporate a nine qubit code to rectify bit-flip errors, \cite{keane2012simplified} that propose a quantum error detection-only circuit on N-qubit systems and two-qubit error correction schemes. 

\subsection{Trapped-ion Qubits}
\paragraph{\textbf{Technological Details}}
Trapped-ion quantum computers use ions as qubits, that are trapped in electromagnetic traps such as the Penning trap that provides confinement in up to two directions or the more widely used RF Paul trap that confines in two or three directions. The ions selected, are usually of alkaline earth metals, such as $Be^+$, $Mg^+$, and $Ca^+$, or of those used in atomic clocks such as $Al^+$, $In^+$, $Lu^+$. These trapped ions are maintained at low temperatures and in vacuum chambers for long coherence times. Quantum gates are implemented using laser pulses that change the quantum state of the ions, and readout is performed by shining fluorescent light on the ions and measuring the intensity of fluorescence (more information can be found in \cite{bruzewicz2019trapped}).   

\paragraph{\textbf{Feasibility}}
Bacon-Shor logical qubit is implemented on 13 trapped ion qubits in \cite{egan2020fault} for error of up to 0.6\% and $>99\%$ fidelity after error correction. [[7,1,3]] Steanne code is built on 10 qubits in \cite{ryan2021realization} that provided fidelity of up to 93\%. Fault-tolerant parity readout has been incorporated in \cite{hilder2022fault} with 93\% parity measurement fidelity. A non-traditional work includes dissipative processing to incorporate a three-qubit code on trapped ions in \cite{reiter2017dissipative}. Older works such as 
\cite{schindler2011experimental} correct single-qubit errors and improve the fidelity of computation by roughly 1\%.

\subsection{Photonic Qubits}
\paragraph{\textbf{Technological Details}}
A photon is used as a qubit, where the life of the qubit starts from the generation of the photon and ends at the detection. The photon is generated usually via processes like spontaneous parametric down-conversion (SPDC) or spontaneous four-wave mixing (SFWM), where a higher energy pump photon is converted into two lower energy daughter photons, and the detection of one of the photons (heralding photon) indicates the presence of another photon (heralded photon). Cryogenic methods like superconducting nanowire single-photon detectors (SNSPD) are used to detect photons with up to $\geq 95\%$ detection efficiency. The state of photonic qubits is given by different degrees of freedom such as, polarization (vertical: $\ket{0}$, horizontal: $\ket{1}$), and spatial modes (such as Orbital Angular Momentum a.k.a OAM), while quantum gates are implemented using different optical devices such as, beam splitters, and phase shifters (more information can be found in \cite{slussarenko2019photonic}).

\paragraph{\textbf{Feasibility}}
OAM-based photons are error corrected in \cite{zhu2021protecting} that provide up to 20\% fidelity in a noisy channel as opposed to less than 1\% fidelity in the uncorrected scenarios. Silicon-based photonic qubits were error corrected, leading to a 30\% increase in overall qubit fidelity in \cite{vigliar2021error}. Bosonic logical qubits have been proposed in \cite{hu2019quantum} that improve fidelity up to 97\%. Older works such as, \cite{pittman2005demonstration} demonstrate error correction on a two-qubit system and achieve up to 98\% fidelity and \cite{kerckhoff2010designing} implements photonic quantum memories to perform quantum error correction that provides around 95\% fidelity.



\section{Challenges and Future Direction} \label{qec_future}


Quantum error correction encounters various hurdles that require resolution to facilitate a fault-tolerant, large-scale quantum computer. Addressing these challenges is critical to executing complex computations with a low error rate. Some of the prevalent challenges include:
\begin{itemize}
    \item \textit{Scalability:} To implement QEC, it is necessary to employ a significant number of qubits to encode a small number of logical qubits. Hence, scaling towards a large number of logical qubits is a substantial challenge.
    \item \textit{Complexity of error decoding algorithms:} Sometimes the complexity of decoding algorithms can be remarkably high, making it challenging to perform real-time QEC.
    \item \textit{Resource overhead:} Error correction necessitates a substantial amount of physical overhead, including gates, qubits, and time. As the code scales towards a higher number of qubits, the overhead also intensifies. Therefore, optimizing the resource overhead is a challenging task.
    \item \textit{Fault-tolerant gates:} The implementation of fault-tolerant quantum gates that can function reliably in the presence of errors is one of the major challenges. While some ECCs permit the use of transversal gates, they do not cover all the gates required for universal quantum computation. \cite{aliferis2005quantum}.
    \item \textit{Noise modeling and error characterization:} Although the development and comprehension of precise noise models are paramount for ensuring the optimal performance of error-correcting codes, it remains a challenging task in quantum systems.
\end{itemize}

Future research in this area may focus on developing more efficient error-correcting codes, enhancing the complexity of decoding algorithms, and refining hardware designs to achieve fault tolerance and improved performance. Hybrid codes that amalgamate the most effective aspects of various types of error-correcting codes are also being investigated by researchers \cite{endo2021hybrid}. Additionally, machine learning is being employed to optimize the performance and scalability of error correction codes \cite{nautrup2019optimizing}. As quantum computing advances, addressing quantum error correction challenges becomes crucial for achieving fault-tolerant computation.
\section{Conclusion} \label{qec_conclusion}

In this paper, we provide a simplified overview of quantum error correction aimed at researchers who may not possess relevant knowledge in quantum physics and mathematics. We begin by discussing classical error correction and its analogous application in the quantum domain, specifically through the use of a $3-qubit$ error correction code. We emphasize that every error-correcting code comprises three components: detection, deduction, and correction. Subsequently, we delve into the stabilizer formalism, examining the necessary properties of a stabilizer to effectively detect quantum errors. Following this, we explore the requirements and general mathematical framework of an encoding circuit before transitioning to topological codes. We discuss both toric and surface codes, comparing their differences despite their shared lattice-based structures for quantum error correction. Lastly, we address current qubit technologies and their application in quantum error-correcting codes. Upon completing this article, readers should possess a fundamental understanding of the workings of quantum error correction, its current state, and potential future developments.

\bibliographystyle{IEEEtran}
\bibliography{references}

\end{document}